\newif\ifsingle

\newif\ifFullVersion
\FullVersiontrue 

\ifsingle
\documentclass[12pt,draftclsnofoot, onecolumn]{IEEEtran}		
\else		
\documentclass[10pt,final, twocolumn]{IEEEtran}
\fi

\usepackage{times}
\usepackage{amsmath,dsfont}
\usepackage{amssymb,amsthm}
\usepackage{epsfig,verbatim}
\usepackage{subcaption}
\usepackage{setspace}
\usepackage{color}
\usepackage{cite}
\usepackage{epstopdf}
\usepackage{graphics}
\usepackage{accents}
\usepackage{acronym}
\usepackage{booktabs}
\usepackage{mathtools} 
\usepackage{enumitem}
\usepackage{multirow}
\usepackage{dblfloatfix}
\usepackage[bookmarks,colorlinks]{hyperref}
\usepackage{gensymb}
\usepackage[flushleft]{threeparttable}
\usepackage{bm}
\usepackage[ruled,linesnumbered]{algorithm2e}  
\usepackage{tikz}
\usetikzlibrary{arrows.meta,positioning,calc,fit}

\SetKwInput{KwData}{\textbf{Init}} 
\let\oldnl\nl
\newcommand{\nonl}{\renewcommand{\nl}{\let\nl\oldnl}}

\newcommand{\myVec}[1]{{\boldsymbol{#1}}}
\newcommand{\myMat}[1]{{\boldsymbol{#1}}}
\newcommand{\mySet}[1]{\mathcal{#1}}
	
\newcommand{\name}{WiMamba}

\ifsingle

\else

\fi 

 \usepackage[all=normal,paragraphs=tight,floats=normal,mathspacing=normal,wordspacing=tight,charwidths=tight,mathdisplays=normal,leading=normal]{savetrees}

\acrodef{adc}[ADC]{analog-to-digital convertor}
\acrodef{cs}[CS]{compressed sensing}
\acrodef{dtft}[DTFT]{discrete-time Fourier transform}
\acrodef{dnn}[DNN]{deep neural network} 
\acrodef{mc}[MC]{monte carlo}
\acrodef{csi}[CSI]{channel state information}
\acrodef{bpsk}[BPSK]{binary phase shift keying}
\acrodef{qpsk}[QPSK]{quadrature phase shift keying}
\acrodef{map}[MAP]{maximum a-posteriori probability}
\acrodef{snr}[SNR]{signal-to-noise ratio}
\acrodef{bs}[BS]{base station} 
\acrodef{iot}[IOT]{Internet of Things}
\acrodef{mmtc}[mMTC]{massive machine-type communications}
\acrodef{gan}[GAN]{Generative Adversarial Network}
\acrodef{embb}[eMBB]{enhanced Mobile Broadband}
\acrodef{mimo}[MIMO]{multiple-input multiple-output}
\acrodef{siso}[SISO]{single-input single-output}
\acrodef{mse}[MSE]{mean-squared error}
\acrodef{pdf}[PDF]{probability density function}
\acrodef{rv}[RV]{random variable}
\acrodef{ml}[ML]{machine learning}
\acrodef{fec}[FEC]{forward error correction}
\acrodef{rs}[RS]{Reed-Solomon}
\acrodef{ar}[AR]{augmented reality}
\acrodef{vr}[VR]{virtual reality}
\acrodef{lti}[LTI]{linear time-invariant}
\acrodef{wss}[WSS]{wide-sense stationary}
\acrodef{psd}[PSD]{power spectral density}
\acrodef{ser}[SER]{symbol error rate} 
\acrodef{ber}[BER]{bit error rate} 
\acrodef{gd}[GD]{gradient descent}
\acrodef{sgd}[SGD]{stochastic gradient descent} 
\acrodef{isi}[ISI]{intersymbol interference}  
\acrodef{awgn}[AWGN]{additive zero-mean white real Gaussian noise} 
\acrodef{ut}[UT]{user terminal} 
\acrodef{mmw}[mmWave]{millimeter wave}
\acrodef{ghz}[GHz]{gigahertz}
\acrodef{noma}[NOMA]{non-orthogonal multiple access}
\acrodef{mac}[MAC]{mulitple access channel}
\acrodef{fl}[FL]{Federated learning}
\acrodef{lstm}[LSTM]{long short-term memory}
\acrodef{maml}[MAML]{model-agnostic meta-learning}
\acrodef{sic}[SIC]{soft interference cancellation}
\acrodef{pmf}[PMF]{probability mass function}
\acrodef{urllc}[URLLC]{ultra-reliable and low-latency communication}
\acrodef{sova}[SOVA]{soft-output Viterbi algorithm}
\acrodef{wbp}[WBP]{weighted belief propagation}
\acrodef{ecc}[ECC]{error-correction codes}
\acrodef{crc}[CRC]{cyclic redundancy check}
\acrodef{scl}[SCL]{successive list cancellation}
\acrodef{bp}[BP]{belief propagation}
\acrodef{mle}[ML]{maximum likelihood}
\acrodef{sota}[SOTA]{state of the art}
\acrodef{irs}[IRS]{intelligent reconfigurable surface}
\acrodef{ssm}[SSM]{state-space model}
\acrodef{bcjr}[BCJR]{Bahl-Cocke-Jelinek-Raviv}
\acrodef{ai}[AI]{artificial intelligence}
\acrodef{nlp}[NLP]{natural language processing}
\acrodef{rnn}[RNN]{recurrent neural network}
\acrodef{gnn}[GNN]{graph neural network}
\acrodef{cnn}[CNN]{Convolutional Neural Network}
\acrodef{gpu}[GPU]{graphics processing unit}
\acrodef{lwm}[LWM]{large wireless model}
\acrodef{mlp}[MLP]{multilayer perceptron}
\acrodef{los}[LOS]{line-of-sight} 
\acrodef{nlos}[NLOS]{non-\ac{los}}
\acrodef{nmse}[NMSE]{normalized mean squared error}
\acrodef{mmse}[MMSE]{minimum mean square error}
\acrodef{mlm}[MLM]{masked language modeling}
\acrodef{mcm}[MCM]{masked channel modeling}

\IEEEoverridecommandlockouts 

\title{\name: Linear-Scale Wireless Foundation Model}
\author{  
	\IEEEauthorblockN{Tomer Raviv and Nir Shlezinger} 
	\thanks{
		The authors are with the School of ECE, Ben-Gurion University of the Negev, Beer-Sheva, Israel (e-mail: tomerraviv95@gmail.com; nirshl@bgu.ac.il). The work was supported  by the European Research Council (ERC) under the ERC starting grant nr. 101163973 (FLAIR).}

	\vspace{-0.5cm}
	
}

\begin{document}

\maketitle

	\pagestyle{plain}
	\thispagestyle{plain}
	\begin{abstract}  
Foundation models learn transferable representations, motivating growing interest in their application to wireless systems. Existing wireless foundation models
are predominantly based on transformer architectures, whose quadratic computational and memory complexity can hinder practical deployment for large-scale channels. In this work, we introduce \name, a wireless foundation model built upon the recently proposed Mamba architecture, which replaces attention mechanisms with selective state-space models and enables linear-time sequence modeling. Leveraging this architectural advantage combined with adaptive preprocessing, \name~achieves scalable and low-latency inference while maintaining strong representational expressivity. We further develop a dedicated task-agnostic, self-supervised pre-training framework tailored to wireless channels, resulting in a genuine foundation model that learns transferable channel representations. Evaluations across four downstream tasks demonstrate that \name~matches or outperforms transformer-based wireless foundation models, while offering dramatic  latency and memory reductions.  
\end{abstract}


    \acresetall
	\vspace{-0.2cm}
\section{Introduction}
\label{sec:introduction}

 
Foundation models have emerged as a powerful \ac{ml} paradigm, centered around large-scale task-agnostic models that learn informative representations transferable across multiple downstream tasks~\cite{yuan2023power}. Their remarkable success in natural language processing and computer vision has  enabled rapid adaptation and improved generalization  in data-limited regimes~\cite{awais2025foundation}. These advances lead to a growing interest in developing foundation models for  communications~\cite{gao2026ai}, and particularly for wireless channels~\cite{jiang2025towards}, where a  learned   \ac{csi} representation  can be reused for diverse tasks such as beam prediction, localization, and channel classification~\cite{guo2025large}. 

Existing  wireless foundation models can be broadly categorized into two classes. The first uses large pre-trained language and vision models, such as the GPT~\cite{zheng2025m2beamllm,liu2024llm4cp,sheng2025beam}, LLaMA~\cite{zheng2025large} and DINO~\cite{guo2025lvm4csi} families, and devises pre- and post-processing pipelines that map wireless data into formats compatible with these architectures, possiblely with limited fine-tuning~\cite{chen2025radiollm}. While this approach leverages the expressive power and maturity of general-purpose foundation models, it relies on large models designed and trained for different modalities. This leads to high computational complexity, significant memory requirements, and when combined with the need for elaborate pre- and post-processing, also introduces notable overhead and latency.

The second class develops \emph{dedicated} foundation models  tailored to wireless channels. This can  be achieved by training a model to process wireless data and alter its parameters for each task via hypernetworks~\cite{zheng2025muse} or fine-tuning~\cite{mashaal2025iqfm}. 
Alternatively, recent works~\cite{yang2025wirelessgpt, alikhani2024large,cheraghinia2025foundation,zhou2025spectrumfm,catak2025bert4mimo,guler2025multi,aboulfotouh20256g} have demonstrated that self-supervised pre-training on large-scale \ac{csi} datasets can yield meaningful and transferable channel representations that benefit a wide range of downstream tasks~\cite{gao2026ai}. However, existing dedicated wireless foundation models rely almost exclusively on transformers, which incur substantial inference latency and memory consumption, particularly when processing large channel inputs. This limitation stems from the quadratic scaling of attention mechanisms with respect to the number of input tokens and the maintenance of key--value caches during inference~\cite{xiao2023efficient}. 

Motivated by these observations, this work introduces a foundation model for wireless channels whose complexity and inference latency scale sub-quadratically with the channel dimensions. Our model, termed {\em \name}, leverages the recently proposed \emph{Mamba} architecture~\cite{gu2024mamba}, which has been shown to achieve expressive sequence modeling comparable to transformers while being substantially more efficient in both computation and memory~\cite{qu2024survey}.  Mamba relies on selective \acp{ssm} rather than self-attention mechanisms, resulting in linear complexity with respect to the input sequence length and eliminating the need for key--value caches in inference. While a few recent works have explored Mamba for specific wireless  tasks~\cite{luo2025cpmamba,li2025deep}, to the best of our knowledge, this is the first work to demonstrate its effectiveness as enabling efficient wireless foundation models.

We design \name~as a wireless foundation model built upon a Mamba bidirectional Mamba backbone, accounting by the fact that the antenna–subcarrier ordering  has no inherent temporal direction. To support varying \ac{csi} dimensionaltiy, we introduce an adaptive granularity mechanism that enables seamless operation across multiple  resolutions  without retraining or architectural modification.   A dedicated task-agnostic, self-supervised pre-training framework allows \name~to learn transferable channel embeddings without  task-specific labels. 
\name~is evaluated on four distinct downstream tasks using DeepMIMO~\cite{alkhateeb2019deepmimo}, covering both classification and regression objectives. Our results demonstrate that \name~achieves performance comparable to, and in several cases exceeding, transformer-based wireless foundation models, while yielding dramatic reductions in inference latency and memory consumption.  


The  paper is structured as follows. Section~\ref{sec:sysmodel} presents the system model. We introduce and evaluate \name~in Sections~\ref{sec:proposed_mamba}-\ref{sec:Simulation}, respectively. Section~\ref{sec:conclusions} provides concluding remarks.


	\vspace{-0.2cm}
\section{System Model}
\label{sec:sysmodel}

\subsection{Problem Formulation}
\label{subsec:CFM}
{\bf Foundation Models}: 
We consider {\em channel foundation models} as task-agnostic models that map \ac{csi} realizations into informative latent representations suitable for a wide range of downstream inference tasks. We focus on multicarrier \ac{mimo} channels with $N$ antennas and $M$ subcarriers, represented by a complex-valued matrix
$\myMat{H} \in \mathbb{C}^{N \times M}$. A channel foundation model is defined as a mapping
\begin{equation}
\myVec{x} = f_{\myVec{\theta}}(\myMat{H}),
\end{equation}
where $f_{\myVec{\theta}}(\cdot)$ is a  function with trainable parameters $\myVec{\theta}$ that extracts a latent representation $\myVec{x}\in \mathbb{R}^D$ from the input \ac{csi}. The foundation model learns $\myVec{\theta}$ from a large set of channel realizations, assumed to capture  diverse  channel conditions, denoted
\begin{equation}
\label{eqn:DataPre}
\mySet{D}_{\mathrm{pre}} = \big\{ \myMat{H}^{(i)} \big\}_{i=1}^{|\mySet{D}_{\mathrm{pre}}|}.
\end{equation}

{\bf Inference Pipeline}: 
The effectiveness of a  foundation model is assessed through its ability to support {\em downstream tasks}. In particular, for a  downstream task indexed by $p$, the inference pipeline is provided with: $(i)$ a relatively small labeled dataset
\begin{equation}
\label{eqn:DataTask}
\mySet{D}_{\mathrm{task}}^{p}
= \big\{ \big(\myMat{H}^{(j)}, \myVec{s}_p^{(j)} \big) \big\}_{j=1}^{|\mySet{D}_{\mathrm{task}}^{p}|},
\end{equation}
where $\myVec{s}_p^{(j)}$ denotes the task-$p$ label associated with the channel realization $\myMat{H}^{(j)}$; and $(ii)$  a task-specific loss function $\ell_p(\cdot,\cdot)$. 
The downstream task is addressed by learning a task-specific mapping from the latent representation $\myVec{x}=f_{\myVec{\theta}}(\myMat{H})$ to the desired output $\myVec{s}_p$, using  the limited data \eqref{eqn:DataTask}. The utility of the foundation model for task $p$ is  quantified by the accuracy with which such a mapping can be learned, as measured by  $\ell_p$.

{\bf Requirements}: 
In this work, we focus on channel foundation models designed to satisfy the following key requirements:
\begin{enumerate}[label={\em R\arabic*}]
\item \label{itm:agnosticism}{\em Task agnosticism}:  $f_{\myVec{\theta}}(\cdot)$ must support multiple downstream tasks that are not known during its design or pre-training.
\item \label{itm:Scalability}{\em Scalability}:   the foundation model should operate on \ac{csi} of varying dimensions  without architectural modifications.
\item \label{itm:Latency}\emph{Latency}: The  model should enable low-latency inference with a limited memory footprint for all channel dimensions. 
\end{enumerate}

\vspace{-0.1cm}
\subsection{Mamba and Selective State-Space Models}
\label{subsec:Mamba}

\begin{figure}
    \centering
    \includegraphics[width=\linewidth]{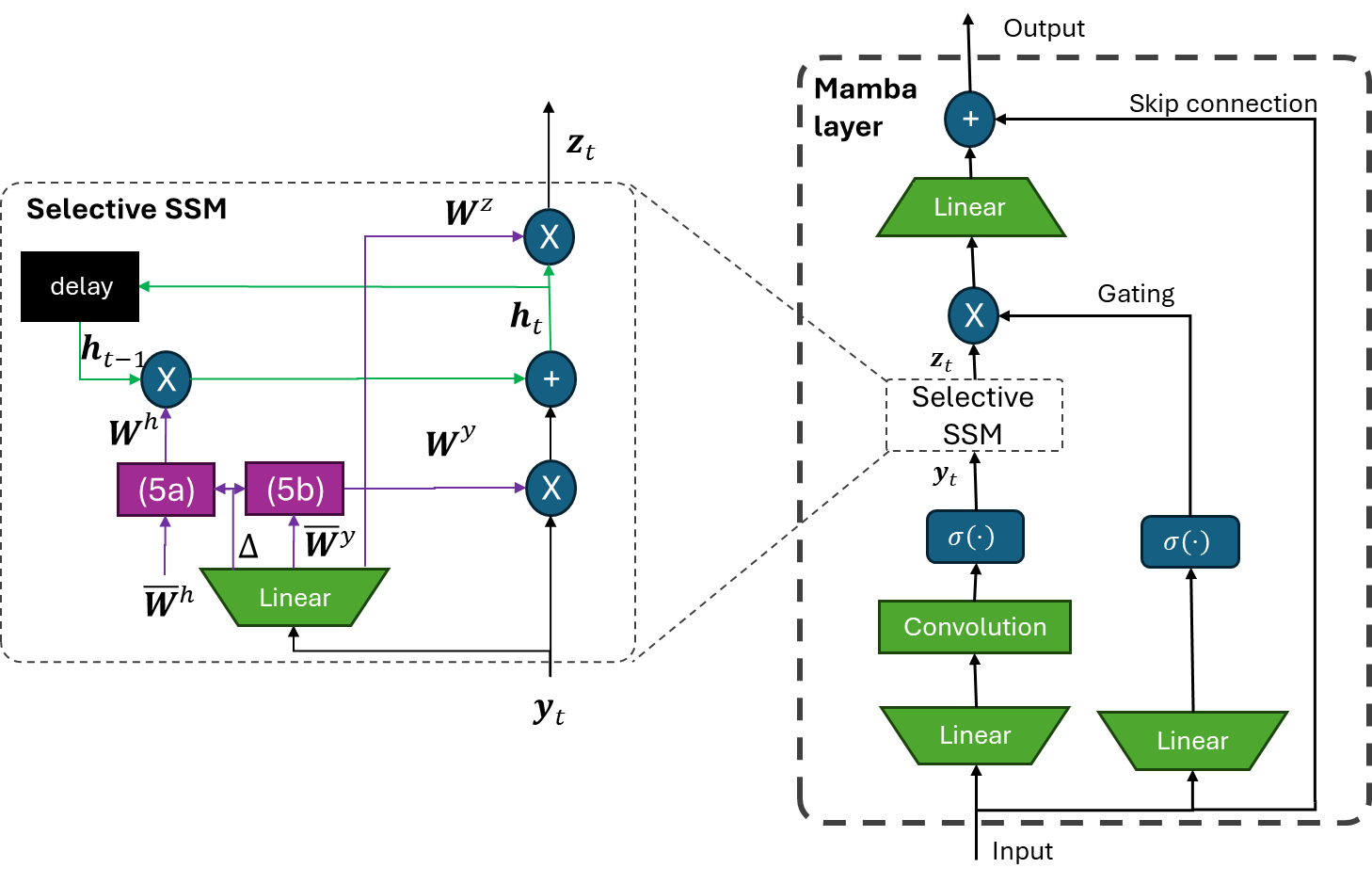}
    \caption{Mamba layer and selective \ac{ssm} illustration. The activation $\sigma(\cdot)$ is typically a sigmoid linear unit.}
    \label{fig:Mamba}
\end{figure}

{\bf Mamba}:
A key ingredient in our design is the recent \emph{Mamba} architecture~\cite{gu2024mamba}, whose core innovation lies in the use of {\em selective \acp{ssm}} for efficient sequence modeling~\cite{gu2021efficiently}. 
At a high level, a Mamba layer processes an input sequence through a stack of a linear projection layer, followed by a lightweight convolutional operation, and a selective \ac{ssm} module. The output of the \ac{ssm} is then gated, projected back to the feature dimension, and combined with a skip connection, as illustrated in Fig.~\ref{fig:Mamba}.

{\bf Selective \acp{ssm}}:
\acp{ssm}-type \ac{ml} modules  parameterize the mapping from an input sequence $\{\myVec{y}_t\}$ into an output sequence $\{{\myVec{z}}_t\}$ through a latent state sequence $\{\myVec{h}_t\}$~\cite{gu2021combining}. \acp{ssm} can be viewed as a form of  \acp{rnn} which operate via evolution and readout equations given by
\begin{align}
\label{eqn:SSM}
\myVec{h}_t &= \myMat{W}^y \myVec{y}_t + \myMat{W}^h \myVec{h}_{t-1}, \qquad 
{\myVec{z}}_t = \myMat{W}^z \myVec{h}_t,
\end{align}
where $\myMat{W}^y$, $\myMat{W}^h$, and $\myMat{W}^z$ are learnable parameters, and the latent state $\myVec{h}_t$ serves as a compact memory that aggregates information from past inputs. In {\em selective} \acp{ssm}~\cite{gu2021efficiently},   $\myMat{W}^y$, $\myMat{W}^h$, and $\myMat{W}^z$ are generated through learned functions of the input $\myVec{y}_t$, effectively resulting in a time-varying and nonlinear \ac{ssm}. Specifically, the state evolution in \eqref{eqn:SSM} is parameterized as
\begin{subequations}
\label{eqn:Mamba}
\begin{align}
\label{eqn:MambaA}
\myMat{W}^h &= \exp\!\left(\Delta \cdot \bar{\myMat{W}}^h\right), \\
\label{eqn:MambaB}
\myMat{W}^y &= \left(\Delta \cdot \bar{\myMat{W}}^h\right)^{-1}
\left(\myMat{W}^h  - \myMat{I}\right)
\Delta \cdot \bar{\myMat{W}}^y,
\end{align}
\end{subequations}
where $\bar{\myMat{W}}^h$ and $\bar{\myMat{W}}^y$ represent continuous-time state transition and input matrices, respectively. Selective \acp{ssm} use a dedicated neural  layer to map $\myVec{y}_t$ into the discretization step $\Delta$, the continuous-time input matrix $\bar{\myMat{W}}^y$, and the output mapping $\myMat{W}^z$. These quantities are then substituted into \eqref{eqn:Mamba} together with the learned $\bar{\myMat{W}}^h$, yielding an input-adaptive state evolution.

{\bf Gains}:
Unlike transformers, which rely on attention mechanisms that correlate all inputs with one another, Mamba utilizes \ac{ssm} representations that enable linear-time processing with respect to the sequence length, while retaining strong expressive power. This property makes Mamba particularly appealing for large-scale wireless channel representations, where both latency and memory efficiency are critical (\ref{itm:Latency}), as proposed next.

	\vspace{-0.2cm}
\section{\name~Channel Foundation Model}
\label{sec:proposed_mamba}

\vspace{-0.1cm}
\subsection{Architecture}
\label{ssec:Architecture}

\name~maps an input \ac{csi} realization $\myMat{H}$ into a latent representation $f_{\myVec{\theta}}(\myMat{H})$ that is  reusable across various downstream inference tasks. Unlike natural language or temporal data, wireless channel matrices lack an inherent causal ordering, while also requiring low-latency processing under strict memory constraints. \name~is therefore designed to accommodate these characteristics by combining two key aspects:  
$(i)$ an {\em adaptive-resolution tokenization and embedding mechanism} that enables flexible control over the sequence length and computational cost; and  
$(ii)$ a {\em bidirectional Mamba backbone} tailored to the non-causal structure of wireless channel matrices.  
We next describe how an input channel is converted into a token sequence and subsequently processed by the proposed backbone.

\subsubsection{Data Preprocessing and Adaptive Tokenization}
\label{subsubsec:preprocessing}
\name~adopts a tokenization procedure that converts the input \ac{csi} matrix into a sequence of fixed-dimensional real-valued tokens. The procedure is inspired by preprocessing pipelines used in existing channel foundation models, e.g.,~\cite{alikhani2024large}, while explicitly enabling adaptive control over the token resolution.  

Specifically, we support a set of $E$ possible token embedding dimension $\{L_e\}_{e=1}^E$, that determine the granularity at which the channel is represented (with the granularity level $e$ treated as additional input). Given an input channel $\myMat{H} \in \mathbb{C}^{N \times M}$, we separate its real and imaginary components and vectorize them. The total number of real-valued channel coefficients is thus $2NM$, and the resulting number of tokens for granularity $e$ is
$T_e = \big\lceil \frac{2NM}{L_e} \big\rceil$.
The token sequence $\{\myVec{t}_1, \ldots, \myVec{t}_{T_e}\}$, with $\myVec{t}_i \in \mathbb{R}^{L_e}$, is obtained by stacking the vectorized real and imaginary parts of $\myMat{H}$,
with zero-padding applied if needed to complete the final token.  Following~\cite{devlin2019bert}, a learnable class token $\myVec{t}^e_{\mathrm{cls}} \in \mathbb{R}^{L_e}$ is prepended to the token sequence to facilitate aggregating global channel information. 
The architecture is designed to support different choices of $L_e$, which balances the resolution-complexity tradeoff of the representation: smaller values of $L_e$ yield longer token sequences that capture finer-grained spatial-frequency structure, whereas larger values reduce the sequence length and processing latency at the expense of representational detail.

\subsubsection{Bidirectional Mamba Backbone}
\label{subsubsec:backbone}

Before being processed, each token is projected into the $D$-dimension embedding space. Specifically, all tokens, including the class token, are mapped from $\mathbb{R}^{L_e}$ to $\mathbb{R}^{D}$ using a shared learnable affine embedding,
\begin{equation}
\label{eqn:Emb}
\myVec{y}_i = \myMat{W}^e_{\mathrm{emb}} \myVec{t}_i + \myVec{b}^e_{\mathrm{emb}},
\qquad
i \in \{\mathrm{cls},1,\ldots,T_e\},
\end{equation}
where $\myMat{W}^e_{\mathrm{emb}} \in \mathbb{R}^{D \times L_e}$ and $\myVec{b}^e_{\mathrm{emb}} \in \mathbb{R}^{D}$.

Since the token sequence corresponds to antenna-subcarrier groupings rather than a temporal evolution, imposing a causal ordering as in standard Mamba may introduce an undesirable inductive bias. Accordingly, the embedded token sequence is then processed by a backbone composed of $Q$  \emph{bidirectional} Mamba layers, drawing inspiration from vision foundation models~\cite{zhu2024vision}. Specifically, each layer employs a forward-directional Mamba and a backward-directional one, each operating with embedding dimension $D/2$. 
Specifically, by dividing each embedding vector into two subvectors via $\myVec{y}_i = [\myVec{y}_i^{\rm F} || \myVec{y}_i^{\rm B}]$ with $\myVec{y}_i^{\rm F},\myVec{y}_i^{\rm B} \in \mathbb{R}^{D/2}$, 
the forward Mamba (with trainable parameters $\myVec{\theta}^{\rm F}$) processes $\myVec{y}^{\rm F}_{\rm cls}, \myVec{y}_1^{\rm F},\ldots, \myVec{y}_{T_e}^{\rm F}$, while the backward Mamba (with trainable parameters $\myVec{\theta}^{\rm B}$) processes the reverse sequence $\myVec{y}_{T_e}^{\rm B}, \myVec{y}_{T_e-1}^{\rm B},\ldots, \myVec{y}_{1}^{\rm B},\myVec{y}_{\rm cls}^{\rm B}$.
The outputs of the two streams are concatenated to form the final representation 
\begin{equation}
\{\myVec{x}_{\mathrm{cls}}, \myVec{x}_1, \ldots, \myVec{x}_{T_e}\},
\qquad
\myVec{x}_i \in \mathbb{R}^{D}.
\label{eqn:EmbOut}
\end{equation}


\begin{figure}
    \centering
    \includegraphics[width=\linewidth]{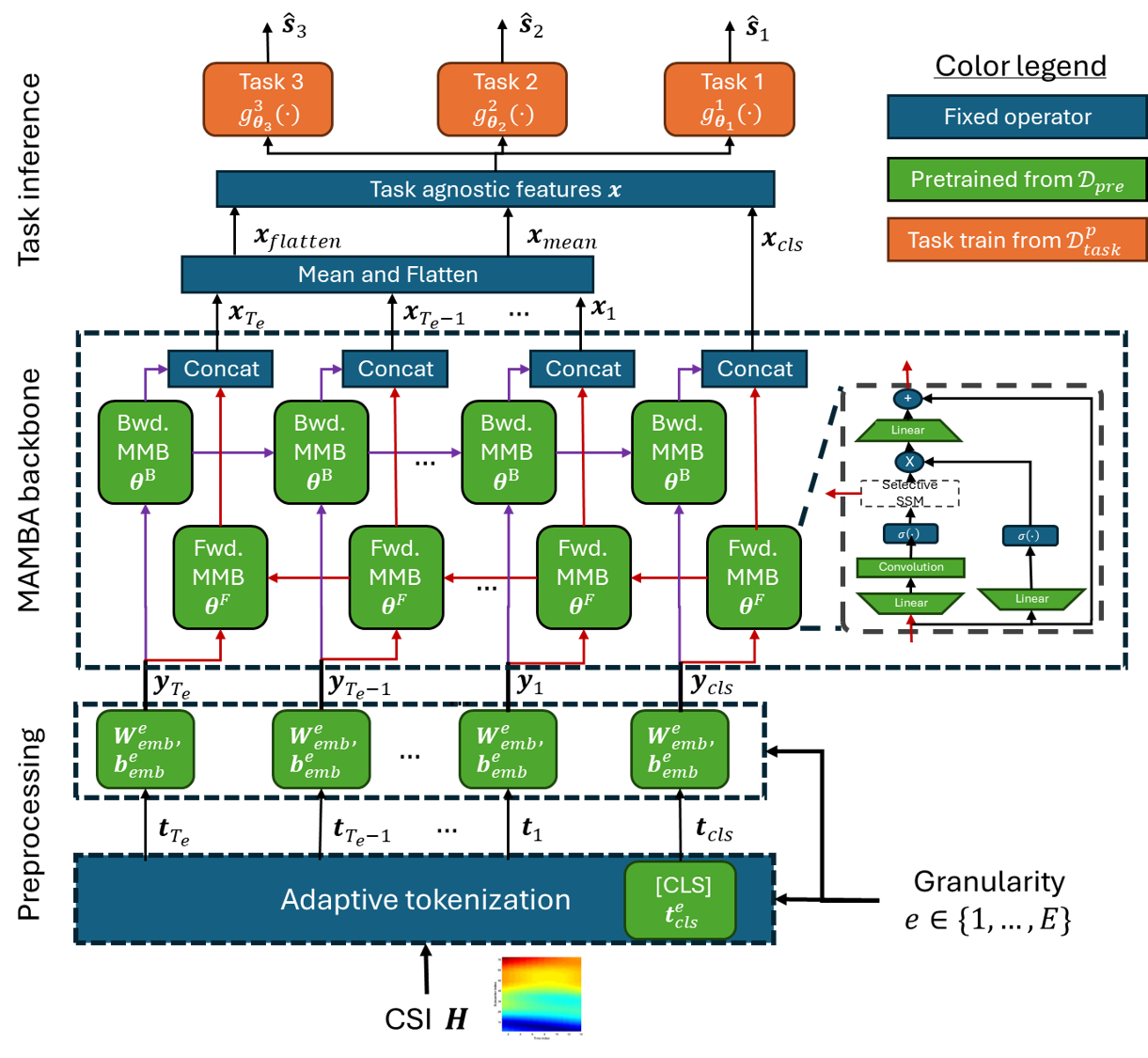}
    \caption{WiMamba  with $Q=1$ forward Mamba (Fwd. MMB) and backward Mamba (Bwd. MMB) layers, and three task heads}
    \label{fig:WiMAMBA}
\end{figure}

\vspace{-0.4cm}
\subsection{Pretraining}
\label{ssec:Pretraining}
We pretrain \name~in a task-agnostic and self-supervised manner using the unlabeled \ac{csi} realizations  \eqref{eqn:DataPre}. Pretraining tunes the trainable parameters $\myVec{\theta} =\{\myVec{\theta}^{\rm F}, \myVec{\theta}^{\rm B}, \{\myMat{W}^e_{\rm emb}, \myVec{b}^e_{\rm emb}, \myVec{t}^e_{\rm cls}\}_{e=1}^E\}$ to have \name~produce contextualized channel representations that capture the underlying structure of wireless channels and can be transferred to diverse downstream tasks with limited labeled data. We follow the masked language modelling practice~\cite{devlin2019bert}, 
along with a dedicated sampling mechanism designed to learn resolution-adaptive representations.

Given a  realization $\myMat{H}$ drawn from the unlabeled dataset \eqref{eqn:DataPre}, we first set a granularity level $e$ uniformly drawn from $\{1,\ldots,E\}$, and apply the preprocessing procedure to obtain a sequence of tokens $\{\myVec{t}_1, \ldots, \myVec{t}_{T_e}\}$. A subset of token indices $\mySet{M}\subset \{1,\ldots,T_e\}$ is then randomly selected for masking. To gurantee that  subsequent masking applied to both the real and its corresponding imaginary components, we set $\mySet{M}$ such that if $i \in \mySet{M}$, and $i \leq T_e/2$ (representing an index of a token taken from ${\rm Re}\{\myMat{H}\}$), then also $i+T_e/2 \in \mySet{M}$ (the corresponding imaginary part). The token $\myVec{t}_i$ for each $i \in \mySet{M}$ is modified according to a predefined masking policy: Inspired by the masked approach of \cite{liu2019roberta},  among the tokens in $\mySet{M}$, $80\%$ are replaced with a ones vector, $10\%$ are replaced with sampled Gaussian vectors, and the remaining $10\%$ are left unchanged.

Pretraining is formulated as a reconstruction problem using an auxiliary linear reconstruction model used only for training, whose goal is  to recover the  masked tokens. Specifically, the masked token sequence is processed by the backbone  to produce
$\{\myVec{x}_{\mathrm{cls}}, \myVec{x}_1, \ldots, \myVec{x}_{T_e}\}$, and these embeddings are mapped into the token dimension using a linear decoder with trainable parameters $\myMat{W}^e_{\rm dec} \in \mathbb{R}^{L_e\times D}$, yielding $\hat{\myVec{t}}_i = \myMat{W}^e_{\rm dec} \myVec{x}_i$. The pretraining loss computed for a token sequence $\{\myVec{t}_i\}$ is  the mean squared error between the reconstructed and original tokens~\cite{liu2019roberta}, 
\begin{equation}
\mathcal{L}_{\mathrm{pre}}(\{\myVec{t}_i\}; \myVec{\theta}, \myMat{W}^e_{\rm dec}) =
\frac{1}{|\mySet{M}|}
\sum_{i \in \mySet{M}}
\left\|
\hat{\myVec{t}}_i - \myVec{t}_i
\right\|_2^2 .
\label{eqn:LossPre}
\end{equation}
The self-supervised objective in \eqref{eqn:LossPre} encourages \name~to learn rich and transferable representations of wireless channels. 

\vspace{-0.2cm}
\subsection{Inference Pipeline}
\label{ssec:Inference} 

A foundation model is  assessed through its ability to support a diverse set of \emph{downstream tasks}. 
The inference pipeline is thus comprised of two stages: $(i)$ the formulation of the new task through a (relatively small) dataset $\mySet{D}_{\rm task}^p$  \eqref{eqn:DataTask} and a loss measure $\ell_p$ (with $p$ being the task index); and $(ii)$ actual task inference applied to a new \ac{csi}. The granularity level $e$ is assumed to be fixed  in inference, based on predefined latency and compute constraints. Therefore, in the following we omit the index $e$ for brevity. 

\subsubsection{Infering Task $p$}
 Given a new task with a channel realization $\myMat{H}$, we apply \name~as described in Subsection~\ref{ssec:Architecture} to  obtain the contextualized embeddings in \eqref{eqn:EmbOut}. The translation of the embeddings $\{\myVec{x}_{\mathrm{cls}},\myVec{x}_1, \ldots, \myVec{x}_T\}$ of channel $\myMat{H}$ into its estimate for the $p$th task, denoted $\hat{\myVec{s}}^p$, is done by mapping the embeddings into a feature vector $\myVec{x}$, and processing it using a dedicated \ac{ml} model trained for task $p$ (referred to as $p$th head).

 The feature vector $\myVec{x}$ is a task-dependent latent representation of $\myMat{H}$ that serves as input to downstream inference heads. It accomodates three types of representations, each capturing a different level of abstraction:
$(i)$ \textbf{Class embedding}, given by the class token $\myVec{x}_{\mathrm{cls}}$, which aggregates information from the entire channel realization into a compact and informative vector; 
$(ii)$ \textbf{Flattened patch embeddings}, obtained as $\myVec{x}_{\mathrm{flatten}}
= \mathrm{vec}\!\left(
\myVec{x}_1, \ldots 
\myVec{x}_T 
\right)$ (zero-padded to length $2NM$), which preserves localized spatial–frequency information across the channel matrix; 
and $(iii)$ \textbf{Mean-pooled patch embeddings}, computed as
$\myVec{x}_{\mathrm{mean}}
=
\frac{1}{T}
\sum_{i=1}^{T}
\myVec{x}_i$,  capturing the overall channel structure while remaining agnostic to the number of patches.

The representation $\myVec{x}\in\{\myVec{x}_{\mathrm{class}},\myVec{x}_{\mathrm{flatten}},\myVec{x}_{\mathrm{mean}}\}$ is processed by a downstream neural network with parameters $\myVec{\theta}_p$, denoted $g_{\myVec{\theta}_p}^p(\cdot)$. The resulting estimate is given by
    $\hat{\myVec{s}}_p = g_{\myVec{\theta}_p}^p(\myVec{x})$.
The overall pipeline  is illustrated in Fig.~\ref{fig:WiMAMBA}, highlighting how different representations are derived from the pretrained backbone.

\subsubsection{Learning Task $p$}
The inference pipeline uses the pretrained \name~to map a channel $\myMat{H}$ to its representation $\myVec{x}$. The only task-specific component is the  neural network $g_{\myVec{\theta}_p}^p(\cdot)$. Accordingly, the dataset \eqref{eqn:DataTask} is required for training solely $\myVec{\theta}_p$, which is based on the 
empirical risk
\vspace{-0.1cm}
\begin{equation}
\mySet{L}_{\mySet{D}_{\mathrm{task}}^{p}}({\myVec{\theta}_p}) = 
\frac{1}{|\mySet{D}_{\mathrm{task}}^{p}|}
\sum_{j=1}^{|\mySet{D}_{\mathrm{task}}^{p}|}
\ell_p
\!\left(g_{\myVec{\theta}_p}^p(\myVec{x}^{(j)}),
\myVec{s}_p^{(j)}
\right),
\vspace{-0.1cm}
\end{equation}
with $\myVec{x}^{(j)}$ is obtained by pretrained \name~applied to $\myMat{H}^{(j)}$.

\subsubsection{Complexity}
A key consideration in our design of \name~is its inference complexity  how it scales with the channel dimensions. For a given granularity level $e$, the token sequence length is $T_e = \lceil 2NM / L_e \rceil$.  The bidirectional Mamba backbone consists of two parallel Mamba models of $Q$ stacked layers, each operating with linear complexity in the sequence length due to the selective \ac{ssm} formulation. As a result, the overall backbone complexity scales as $\mathcal{O}(Q T_e D) = \mathcal{O}(Q D \lceil 2NM / L_e \rceil)$, which is linear in both $N$ and $M$. Importantly, the granularity parameter $e$ (through $L_e$), enables balancing computational cost, memory usage, and representation fidelity. Unlike transformer-based models, whose complexity grows quadratically with $T_e$~\cite{keles2023computational}, \name~maintains linear scaling even for high-resolution channel representations. 


	\vspace{-0.2cm} 
\section{Numerical Evaluations}
\label{sec:Simulation}
\subsection{Experimental Setup}
{\bf Data:} 
We numerically evaluate  \name\footnote{The source code used in our empirical study, along with the hyperparameters is available at \url{https://github.com/tomerraviv95/mamba-lwm-project}} using the DeepMIMO dataset~\cite{alkhateeb2019deepmimo}. We  construct the pretraining dataset $\mySet{D}_{\rm pre}$ using  channel realizations  with dimensions $16\times16$, $32\times32$, $64\times64$, $128\times128$, and $256\times256$, and employ  adaptive tokenization with sizes $L_e\in\{16, 64\}$, respectively corresponding to \ac{csi}  patch sizes of
$4\times4$ and $8\times8$.

We focus on four downstream tasks accomodated by DeepMIMO: \ac{los} vs. \ac{nlos} classification, beam prediction, channel interpolation, and user localization. 
The loss measures are cross entropy for classification tasks (\ac{los}/\ac{nlos} and beam prediction) or $\ell_2$ loss for channel interpolation and localization. Each downstream task is handled by a dedicated \ac{mlp}-based head operating on the patch embeddings, which are: 
\begin{itemize}
    \item {\em \ac{los}/\ac{nlos} classification} uses a shallow \ac{mlp} with an 8-unit hidden layer to output binary logits.
    \item {\em Beam prediction} employs an MLP with 256- and 128-unit hidden layers to produce logits over 64 beams. 
    \item {\em Channel interpolation} maps each patch embedding through a linear layer to a $P\times P\times 2$ spatial patch, followed by deterministic patch reassembly to recover the full channel. 
    \item {\em User localization} applies a three-layer \ac{mlp} with hidden dimensions 64 and 32 to output 2-D coordinates.
\end{itemize}

\begin{figure}
    \centering
    
    \begin{subfigure}{\linewidth}
        \centering
        \includegraphics[width=\linewidth]{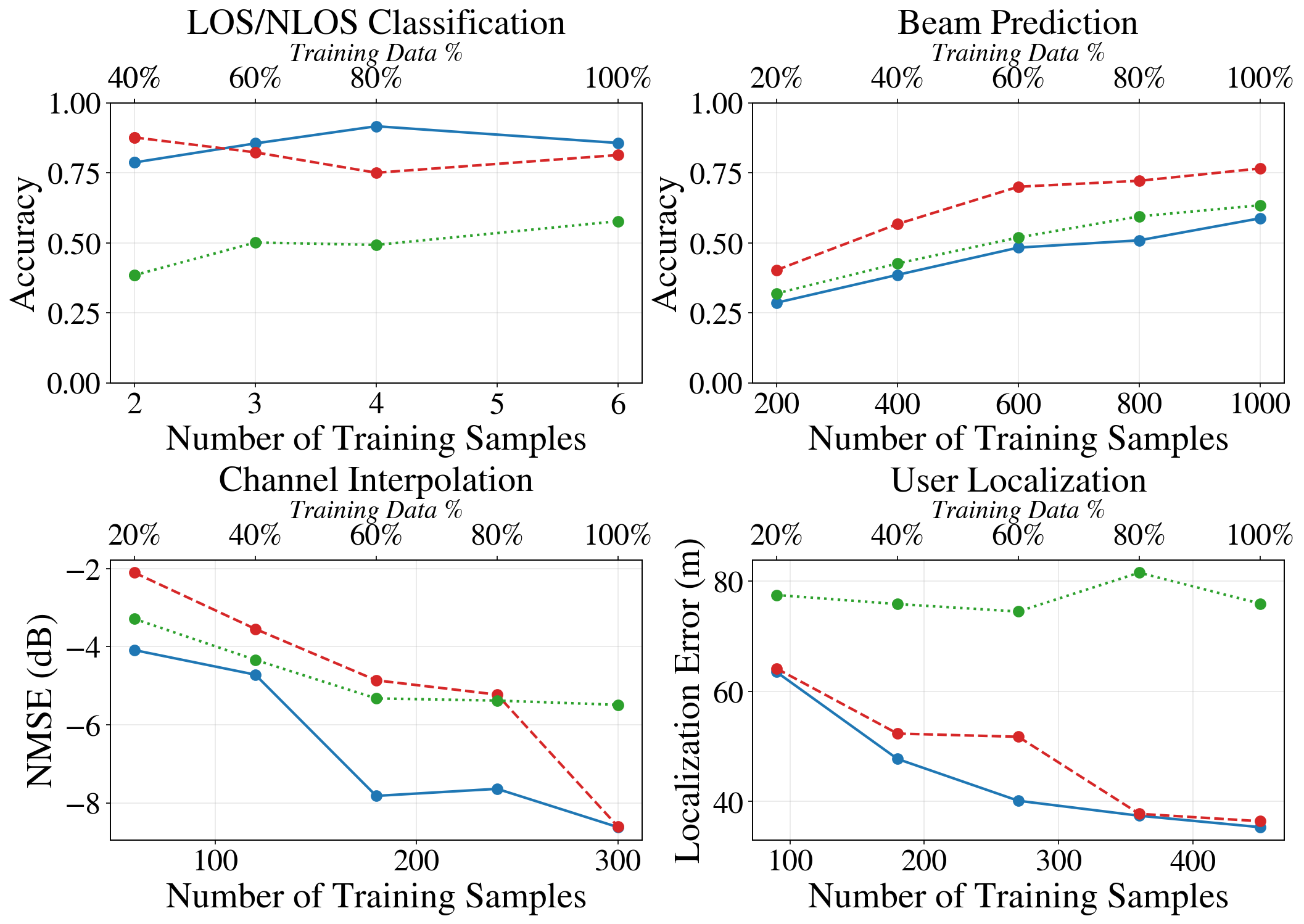}
        \caption{4$\times$4 Patch Size}
    \end{subfigure}
    
    
    
    \vspace{0.2cm}
    
    \begin{subfigure}{\linewidth}
        \centering
        \includegraphics[width=\linewidth]{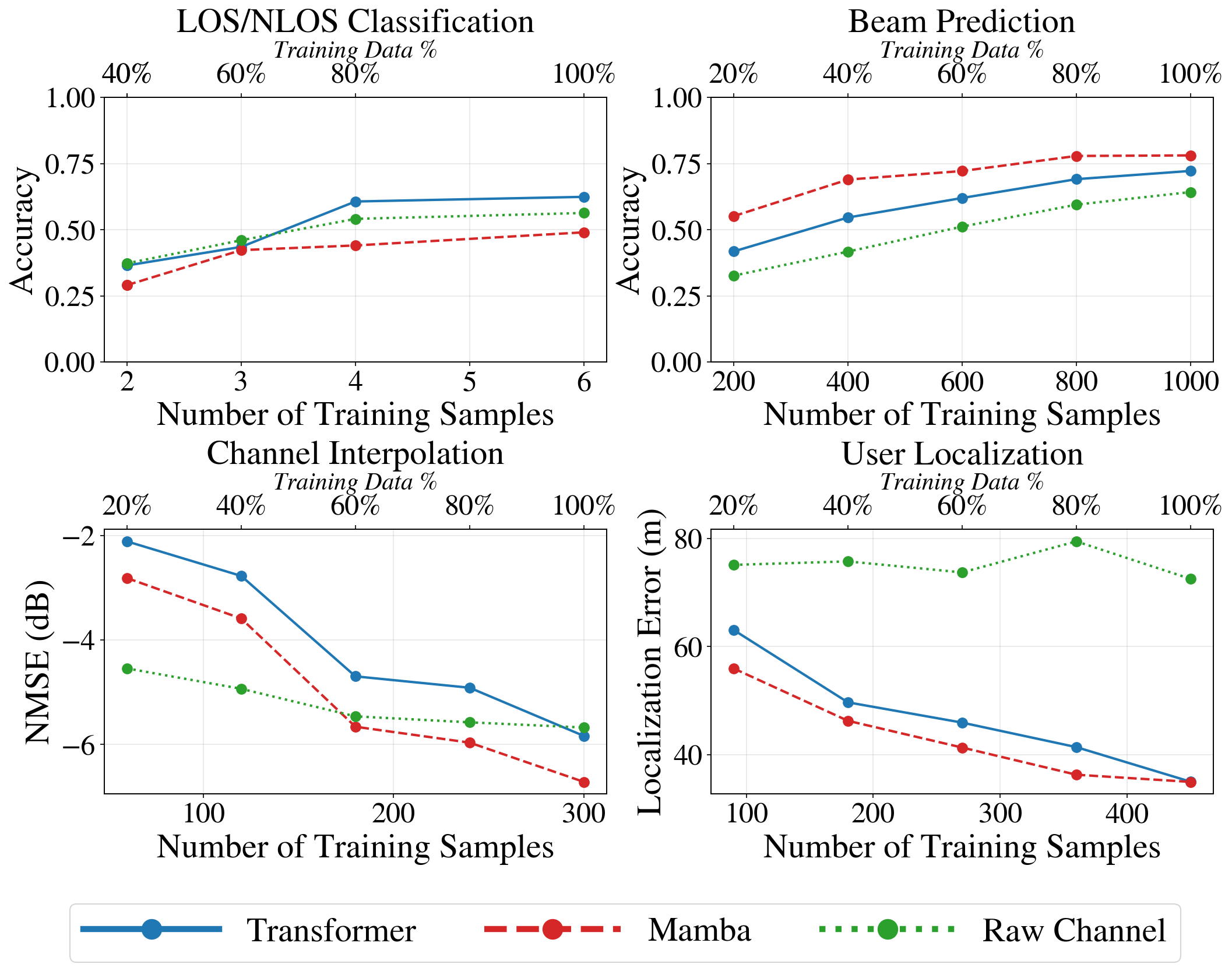}
        \caption{8$\times$8 Patch Size}
    \end{subfigure}
    
    \caption{Task performance  vs. $|\mySet{D}^p_{\rm task}|$ 
    }
    \label{fig:patch_tradeoff}
\end{figure}
\smallskip
{\bf Algorithms:} 
We implement \name~using $Q=12$ bidirectional Mamba layers, with an overall embedding size of $D=128$ (with each directional Mamba processing $D/2=64$ features). The Mamba cells use an  $8\times 1$ state  ($\myVec{h}_t$ in \eqref{eqn:SSM}) with local convolution kernel size of $4$ to capture short-range interactions.  
Our main benchmark is the transformer-based \ac{lwm} 1.1 model of \cite{alikhani2024large}, comprised of $12$ transformer encoder layers with embedding size of $D=128$.  Both foundation models have $2.5\cdot 10^6$ paramteres, and are pre-trained for $100$ epochs using a batch size of $32$ for both training and validation.
Training hyperparameters are selected via empirical trials, and are based on those used in  \cite{alikhani2024large}.

\vspace{-0.1cm}
\subsection{Experimental Results}
\label{subsec:task_performance}
{\bf Task Performance:}
We first evaluate performance on the considered tasks.
Here we examine whether replacing transfomers with \ac{ssm}-based Mamba impacts task-level performance.  
Accordingly, we compare the performance achieved for each task when training the dedicated task head based on the outputs of the foundation models, as well as  when training using the raw channels in $\mySet{D}_{\rm task}^p$  without any learned representation. 

Fig.~\ref{fig:patch_tradeoff} presents the downstream task performance for the Transformer-based \ac{lwm}, \name, and a raw-channel baseline. For each resolution, we vary the number of task-specific training samples and report accuracy for classification tasks and \ac{nmse} or mean absolute error  for regression tasks.
We first observe a consistent improvement in performance as the number of labeled training samples increases, across all tasks and patch sizes. This trend confirms that the representations extracted by both foundation models are effectively leveraged by the lightweight downstream heads.

Comparing architectures, \name~achieves performance that is consistently competitive with the Transformer baseline. For the smallest patches (4$\times$4), where the sequence length is longest, the two models exhibit closely matched performance. 
Minor differences appear across individual tasks and data regimes, but no consistent performance gap emerges in favor of the Transformer.
These results support our key conclusion,  that properly designed \ac{ssm}-based models can lead to comparable representations as that of transformer-based foundation models.


\smallskip
{\bf Latency and Memory Evaluation:}
We next evaluate the efficiency of \name~under varying token granularity, in comparison with the transformer-based \ac{lwm}. Since the patch size  controls the token sequence length $T_e$, smaller patches induce longer sequences and higher computational burden. 

Table~\ref{tab:latency_comparison} reports both the inference latency and the peak \acs{gpu} memory consumption of the two architectures as a function of the patch size, evaluated on the same  NVIDIA RTX~3060. \ac{lwm} exhibits a sharp increase in latency and memory usage as the patch size decreases from $8\times 8$ to $4\times 4$, the latter leading 
to substantially higher inference time and nearly an order-of-magnitude increase in peak memory due the quadratic cost of self-attention with  the sequence length.
In contrast, \name~maintains consistently low latency and memory consumption across all patch sizes. This stable behavior highlights its linear-scaling and the flexibility provied through our adaptive tokenization mechanism. Overall, these results confirm the key complexity advantage of \name. 

\begin{table}
\centering
\scriptsize
\caption{Inference latency and peak \acs{gpu} memory.}
\label{tab:latency_comparison}
\resizebox{\columnwidth}{!}{%
\begin{tabular}{ccccc}
\toprule
\multirow{2}{*}{Patch} & \multicolumn{2}{c}{Transformer} & \multicolumn{2}{c}{Mamba} \\
\cmidrule(lr){2-3} \cmidrule(lr){4-5}
 & Lat.~(ms) & Mem.~(MB) & Lat.~(ms) & Mem.~(MB) \\
\midrule
$4 \times 4$ & 363.59 & 4836.63 & 23.70 & 114.40 \\
$6 \times 6$ &  79.15 & 1002.19 & 16.17 &  75.38 \\
$8 \times 8$ &  29.42 &  355.67 & 16.21 &  62.41 \\
\bottomrule
\end{tabular}%
}
\end{table}

    \vspace{-0.2cm}
\section{Conclusion}
\label{sec:conclusions}

We introduced WiMamba, a wireless foundation model built on a selective \acp{ssm}  tailored to the structural properties of \ac{csi}. Through task-agnostic self-supervised pretraining and an adaptive granularity mechanism, WiMamba learns transferable channel representations that operate seamlessly across multiple patch resolutions without retraining.  WiMamba is shown to match transformer-based baselines in downstream performance, 
while offering notably improved computational scalability and latency. 
	\bibliographystyle{IEEEtran}
	\bibliography{IEEEabrv,refs}

\begin{thebibliography}{10}
\providecommand{\url}[1]{#1}
\csname url@samestyle\endcsname
\providecommand{\newblock}{\relax}
\providecommand{\bibinfo}[2]{#2}
\providecommand{\BIBentrySTDinterwordspacing}{\spaceskip=0pt\relax}
\providecommand{\BIBentryALTinterwordstretchfactor}{4}
\providecommand{\BIBentryALTinterwordspacing}{\spaceskip=\fontdimen2\font plus
\BIBentryALTinterwordstretchfactor\fontdimen3\font minus \fontdimen4\font\relax}
\providecommand{\BIBforeignlanguage}[2]{{%
\expandafter\ifx\csname l@#1\endcsname\relax
\typeout{** WARNING: IEEEtran.bst: No hyphenation pattern has been}%
\typeout{** loaded for the language `#1'. Using the pattern for}%
\typeout{** the default language instead.}%
\else
\language=\csname l@#1\endcsname
\fi
#2}}
\providecommand{\BIBdecl}{\relax}
\BIBdecl

\bibitem{yuan2023power}
Y.~Yuan, ``On the power of foundation models,'' in \emph{International Conference on Machine Learning}.\hskip 1em plus 0.5em minus 0.4em\relax PMLR, 2023, pp. 40\,519--40\,530.

\bibitem{awais2025foundation}
M.~Awais \emph{et~al.}, ``Foundation models defining a new era in vision: a survey and outlook,'' \emph{{IEEE} Trans. Pattern Anal. Mach. Intell.}, vol.~47, no.~4, pp. 2245--2264, 2025.

\bibitem{gao2026ai}
Y.~Gao \emph{et~al.}, ``{AI}-driven channel state information extrapolation for 6{G}: Current situations, challenges and future research,'' \emph{arXiv:2601.00159}, 2026.

\bibitem{jiang2025towards}
J.~Jiang \emph{et~al.}, ``Towards channel foundation models ({CFM}s): Motivations, methodologies and opportunities,'' \emph{arXiv:2507.13637}, 2025.

\bibitem{guo2025large}
J.~Guo, Y.~Cui, S.~Jin, and J.~Zhang, ``Large {AI} models for wireless physical layer,'' \emph{arXiv:2508.02314}, 2025.

\bibitem{zheng2025m2beamllm}
C.~Zheng \emph{et~al.}, ``{M2BeamLLM}: Multimodal sensing-empowered mm{W}ave beam prediction with large language models,'' \emph{arXiv:2506.14532}, 2025.

\bibitem{liu2024llm4cp}
B.~Liu \emph{et~al.}, ``{LLM4CP}: Adapting large language models for channel prediction,'' \emph{J. Commn. Net’}, vol.~9, no.~2, pp. 113--125, 2024.

\bibitem{sheng2025beam}
Y.~Sheng \emph{et~al.}, ``Beam prediction based on large language models,'' \emph{{IEEE} Wireless Commun. Lett.}, vol.~14, no.~5, pp. 1406--1410, 2025.

\bibitem{zheng2025large}
T.~Zheng and L.~Dai, ``Large language model enabled multi-task physical layer network,'' \emph{{IEEE} Trans. Commun.}, vol.~74, pp. 307--321, 2025.

\bibitem{guo2025lvm4csi}
J.~Guo \emph{et~al.}, ``{LVM4CSI}: Enabling direct application of pre-trained large vision models for wireless channel tasks,'' \emph{arXiv:2507.05121}, 2025.

\bibitem{chen2025radiollm}
S.~Chen \emph{et~al.}, ``Radio{LLM}: Introducing large language model into cognitive radio via hybrid prompt and token reprogrammings,'' \emph{arXiv:2501.17888}, 2025.

\bibitem{zheng2025muse}
T.~Zheng \emph{et~al.}, ``{MUSE-FM}: Multi-task environment-aware foundation model for wireless communications,'' \emph{arXiv:2509.01967}, 2025.

\bibitem{mashaal2025iqfm}
O.~Mashaal and H.~Abou-Zeid, ``{IQFM}: A wireless foundational model for {I/Q} streams in {AI}-native {6G},'' \emph{arXiv:2506.06718}, 2025.

\bibitem{yang2025wirelessgpt}
T.~Yang \emph{et~al.}, ``Wireless{GPT}: A generative pre-trained multi-task learning framework for wireless communication,'' \emph{{IEEE} Netw.}, 2025.

\bibitem{alikhani2024large}
S.~Alikhani, G.~Charan, and A.~Alkhateeb, ``Large wireless model ({LWM}): A foundation model for wireless channels,'' \emph{arXiv:2411.08872}, 2024.

\bibitem{cheraghinia2025foundation}
M.~Cheraghinia \emph{et~al.}, ``Foundation model for wireless technology recognition using {IQ} timeseries,'' \emph{arXiv:2505.19390}, 2025.

\bibitem{zhou2025spectrumfm}
F.~Zhou \emph{et~al.}, ``Spectrum{FM}: A foundation model for intelligent spectrum management,'' \emph{arXiv:2505.06256}, 2025.

\bibitem{catak2025bert4mimo}
F.~O. Catak, M.~Kuzlu, and U.~Cali, ``{BERT4MIMO}: A foundation model using bert architecture for massive {MIMO} channel state information prediction,'' \emph{arXiv:2501.01802}, 2025.

\bibitem{guler2025multi}
B.~Guler, G.~Geraci, and H.~Jafarkhani, ``A multi-task foundation model for wireless channel representation using contrastive and masked autoencoder learning,'' \emph{arXiv:2505.09160}, 2025.

\bibitem{aboulfotouh20256g}
A.~Aboulfotouh, E.~Mohammed, and H.~Abou-Zeid, ``{6G WavesFM}: A foundation model for sensing, communication, and localization,'' \emph{arXiv:2504.14100}, 2025.

\bibitem{xiao2023efficient}
G.~Xiao \emph{et~al.}, ``Efficient streaming language models with attention sinks,'' in \emph{International Conference on Learning Representations}, 2024.

\bibitem{gu2024mamba}
A.~Gu and T.~Dao, ``Mamba: Linear-time sequence modeling with selective state spaces,'' in \emph{First conference on language modeling}, 2024.

\bibitem{qu2024survey}
H.~Qu \emph{et~al.}, ``A survey of {M}amba,'' \emph{arXiv:2408.01129}, 2024.

\bibitem{luo2025cpmamba}
S.~Luo, J.~Xie, Y.~Che, J.~Yao, J.~Tian, D.~Feng, and K.~Wu, ``{CPM}amba: Selective state space models for {MIMO} channel prediction in high-mobility environments,'' \emph{arXiv:2512.16315}, 2025.

\bibitem{li2025deep}
Z.~Li, C.~Zheng, J.~Xiao, J.~Wang, G.~Wang, M.~Zeng, and O.~A. Dobre, ``Deep learning based joint channel estimation and positioning for sparse {XL-MIMO OFDM} systems,'' \emph{arXiv:2507.19936}, 2025.

\bibitem{alkhateeb2019deepmimo}
A.~Alkhateeb, ``Deep{MIMO}: A generic deep learning dataset for millimeter wave and massive {MIMO} applications,'' \emph{arXiv:1902.06435}, 2019.

\bibitem{gu2021efficiently}
A.~Gu, K.~Goel, and C.~R{\'e}, ``Efficiently modeling long sequences with structured state spaces,'' \emph{International Conference on Learning Representations}, 2022.

\bibitem{gu2021combining}
A.~Gu \emph{et~al.}, ``Combining recurrent, convolutional, and continuous-time models with linear state space layers,'' \emph{Advances in Neural Information Processing Systems}, 2021.

\bibitem{devlin2019bert}
J.~Devlin, M.-W. Chang, K.~Lee, and K.~Toutanova, ``{BERT}: Pre-training of deep bidirectional transformers for language understanding,'' in \emph{Conference of the North American Chapter of the Association for Computational Linguistics: Human Language Technologies}, 2019, pp. 4171--4186.

\bibitem{zhu2024vision}
L.~Zhu \emph{et~al.}, ``Vision {M}amba: Efficient visual representation learning with bidirectional state space model,'' \emph{arXiv:2401.09417}, 2024.

\bibitem{liu2019roberta}
Y.~Liu \emph{et~al.}, ``{RoBERTa}: A robustly optimized bert pretraining approach,'' \emph{arXiv:1907.11692}, 2019.

\bibitem{keles2023computational}
F.~D. Keles, P.~M. Wijewardena, and C.~Hegde, ``On the computational complexity of self-attention,'' in \emph{International Conference on Algorithmic Learning Theory}.\hskip 1em plus 0.5em minus 0.4em\relax PMLR, 2023, pp. 597--619.

\end{thebibliography}

\end{document}